\documentclass[aps,pra,amsmath, amsfonts, amssymb, superscriptaddress, twocolumn, showpacs]{revtex4}

\usepackage{graphicx}
\usepackage{wasysym}
\usepackage{color}
\usepackage{nicefrac}

\def\a{s}
\def\b{s}
\renewcommand{\vec}[1]{\mathbf{#1}} 
\newcommand{\add}[1]{\if\a\b{{\color{red} #1}}\else{#1}\fi}
\newcommand{\comm}[1]{\if\a\b{{\color{blue} #1}}\else{#1}\fi}

\newcommand{\citeasnoun}[1]{Ref.~\onlinecite{#1}}

\newcommand{\eqnumref}[1]{(\ref{eq:#1})}
\renewcommand{\eqref}[1]{Eq.~\eqnumref{#1}}
\newcommand{\Eqref}[1]{Equation~\eqnumref{#1}}

\newcommand{\secref}[1]{Sec.~\ref{sec:#1}}

\newcommand{\figref}[1]{Fig.~\ref{fig:#1}}

\renewcommand{\Re}{\operatorname{Re}}
\renewcommand{\Im}{\operatorname{Im}}
\newcommand{\mean}{\operatorname{mean}}
\newcommand{\J}{\mathcal{J}}

\begin{document}

\title{General scaling limitations of ground-plane and isolated-object cloaks}

\author{Hila Hashemi}
\affiliation{Department of Mathematics, Massachusetts Institute of Technology, Cambridge, MA 02139}
\author{A. Oskooi}
\affiliation{Department of Electronic Science and Engineering, Kyoto University, Japan}
\author{J.~D.~Joannopoulos}
\affiliation{Department of Physics, Massachusetts Institute of Technology, Cambridge, MA 02139}
\author{Steven~G.~Johnson}
\affiliation{Department of Mathematics, Massachusetts Institute of Technology, Cambridge, MA 02139}

\begin{abstract}
We prove that, for arbitrary three-dimensional transformation-based
invisibility cloaking of an object above a ground plane or of isolated
objects, there are practical constraints that increase with the object
size.  In particular, we show that the cloak thickness must scale
proportional to the thickness of the object being cloaked, assuming
bounded refractive indices, and that absorption discrepancies and
other imperfections must scale inversely with the object
thickness. For isolated objects, we also show that bounded refractive
indices imply a lower bound on the effective cross-section.
\end{abstract}

\maketitle

\section{Introduction}

Invisibility ``cloaking'' refers to the idea of making an object
appear invisible, or at least greatly reducing its scattering
cross-section, by surrounding it with appropriate materials, and has
attracted extensive popular and research interest following a
theoretical proposal by Pendry~\cite{Pendry06}.  Unfortunately,
cloaking of isolated objects turns out to be severely restricted, in
that speed-of-light/causality constraints intrinsically limit perfect
cloaking to an infinitesimal bandwidth~\cite{Pendry06, Miller06}.  An
alternative with no intrinsic bandwidth limitations is ground-plane
cloaking~\cite{Li08}, in which the goal is to make an object sitting
on a reflective surface indistinguishable from the bare surface.  In
this paper, however, we show that \emph{both} forms of cloaking are
subject to practical difficulties that increase as the size of the
cloaked object grows, generalizing a simple one-dimensional argument
that we previously applied to ground-plane cloaking~\cite{Hashemi10}.
We focus most of our attention on the case of ground-plane cloaking,
which seems to be the most practical possibility, but we then show
that a very similar analysis applies to cloaking of isolated objects.
In both cases, our key starting point is the assumption that the
attainable refractive indices are bounded, in which case we show that
the cloak thickness must scale with the object size and hence any
losses per unit volume (including both absorption and scattering from
imperfections) must scale inversely with the object size. It has been
suggested that gain could be used to compensate for absorption loss
(but not other imperfections) in the cloak~\cite{Han11}, but a
corollary of our results is that such compensation must become
increasingly exact as the object diameter increases.

Although there have been several experimental demonstrations of
ground-plane cloaking~\cite{Liu09, Ma09OE, Gabrielli09, Valentine09,Lee09, ErginSt10, Ma10, Zhang11, Chen11, Gharghi11}, as well as theoretical
investigation of several variations on the underlying
idea~\cite{ErginSt10OE, Xu09, Baile10, Landy10}, we previously argued
using a simple one-dimensional model system that the difficulty of
ground-plane cloaking must increase proportional to the thickness of
the object being cloaked~\cite{Hashemi10}. In this paper, we
generalize that simple argument to a rigorous proof for arbitrary
cloaking transformations in three dimensions.  We demonstrate that the
thickness of the cloak must scale proportional to the size of the
object, given bounded material properties.  (Previously, we arrived at
a similar conclusion in one dimension from the delay--bandwidth
product~\cite{Hashemi10}, but here our result is derived independent
of the bandwidth, although the bounds on the indices ultimately depend
on the bandwidth.) From this, if one requires a bounded reduction in
the scattering cross-section, it follows that the loss (due to absorption
or other imperfections) per unit volume must scale inversely with the
object thickness, and we quantify this scaling more precisely in the
case of absorption loss and scattering from disorder.  For a lossy
ambient medium such as a fluid (with an observer close enough to see
the object), the losses of the cloak must asymptotically approach
those of the ambient medium, and defects/roughness that scatter light
must still vanish, so there is still a sensitivity to imperfections.
(On the other hand, ambient fluids have the advantage that it is
easier to index-match them with a solid cloak without resorting to
complicated metamaterial microstructures susceptible to manufacturing
imperfections. This helped recent authors, using natural birefringent
materials, to demonstrate cloaking effects at visible wavelengths for
cm-scale structures~\cite{Zhang11,Chen11}.)  In addition to scattering
and loss, systematic imperfections (such as an overall shift in the
indices or an overall neglect of anisotropy in favor of approximate
isotropic materials~\cite{Li08}) must also vanish inversely with
object thickness, since such systematic errors produce a worst-case
phase shift in the reflected field proportional to the imperfection
and the thickness of the cloak (the path length, which scales with the
object).  (For oblique angles of incidence, such phase shifts can
cause a lateral shift in the reflected beam~\cite{Baile10}, analogous
to a Goos-H{\"a}nchen shift.)

Cloaking of isolated objects, on the other hand, is already subject to
severe bandwidth restrictions: perfect cloaking in vacuum over a
nonzero bandwidth would imply rays traveling at greater than the speed
of light around the object~\cite{Pendry06}, which can be interpreted
as a causality violation~\cite{Miller06}.  Nevertheless, the
isolated-cloaking problem is of considerable fundamental theoretical
interest~\cite{Leonhardt06NJP, Schurig06OE, Cummer06, Qiu09, Chen07PRB, Kwon08,Jiang08, Kante08, Ma09, Yan07, Ruan07,Huang07, Zhang08APL, Cai07NP, Zolla07,Tucker05, Nicolet08, Chen08JAP, Leonhardt09, Baile09, Argyropoulos10, Baile10OE, Han10, Han11}, and
several groups have demonstrated single-frequency cloaking of small
objects in experiments~\cite{Schurig06, Smolyaninov08, Kante09,
  Liu09APL}.  Although several theoretical simulations included
absorption loss~\cite{Li09, Chen07PRL, Cummer06, Baile10OE, Cai07NP, Han11, Kante08, Jiang08}, the first work we are aware of to suggest a
tradeoff between absorption tolerance and object size was
\citeasnoun{Baile09}.  Based on these numerical
experiments~\cite{Baile09} and on comparison with ground-plane
cloaking, we suggested~\cite{Hashemi10} that even single-frequency
isolated-object cloaking must become increasingly difficult as the cloaked
object becomes bigger, and in this paper we are able to prove that
result analytically. In particular, assuming that the attainable
refractive indices are bounded above, we show that the cloak thickness
must scale proportional to the object diameter and any cloak losses
(absorption or imperfections) must scale inversely with diameter.
(Experimentally, the group to claim isolated cloaking of an
object more than wavelength-scale in diameter~\cite{Smolyaninov09PRL}
did not do so in vacuum, but rather within a parallel-plate waveguide
system free of complex microstructures and hence with very low
intrinsic losses.)  Another limitation on isolated object cloaking is
that the singularity of the cloaking transformation (which maps an
object to a single point) corresponds to very extreme material
responses (e.g. vanishing effective indices) at the inner surface of a
perfect cloak~\cite{Baile}.  Here, independent of our results on
losses, we show that if the attainable refractive indices are bounded
below, then the cloak is necessarily imperfect: it reduces the object
cross section by a bounded fraction, even for otherwise lossless and
perfect materials.  (Previous authors showed that the bounded
reduction in the cross section obtained from a non-singular cloaking
transformation could be partially defeated by resonant inclusions in
the object~\cite{KohnOn10}, but this problem seems avoidable by a
cloak with a reflective inner surface that masks the nature of the
cloaked object.)  Both of these diameter-scaling limitations are
apparent if one looks at explicit examples of cloaking
transformations, such as Pendry's original linear
scaling~\cite{Pendry06}, but our results differ from such observations
in that they hold in general for any arbitrary transformation,
including transformations that are not spherically symmetrical.

\section{Ground-plane cloaking}

\begin{figure}[t]                                          
 \centering \includegraphics[width=0.5\textwidth]{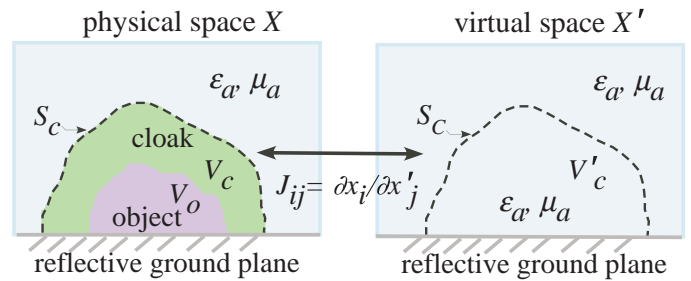}
 \caption{Schematic of a general cloaking problem: an object in a
   volume $V_o$ sitting on a reflective ground is cloaked by choosing
   the materials $\varepsilon$ and $\mu$ in a surrounding volume $V_c$
   to mimic a coordinate transformation, with Jacobian $\J$, mapping
   the physical space $X$ to a virtual space $X'$ in which the object
   is mapped into the ground and $V_c$ is mapped into the entire
   $V_c'=V_c \cup V_o$ volume with the homogeneous ambient-space
   properties $\varepsilon_a$ and $\mu_a$. $S_c$ denotes the outer
   surface of the cloak (identical in $X$ and $X'$).}
 \label{fig:cloak-transf}
\end{figure} 

Consider an object in a volume $V_o$ on a reflective ground plane,
surrounded by a homogeneous isotropic ambient medium with permittivity
$\varepsilon_a$ and permeability $\mu_a$.  This object is cloaked by
choosing the materials $\varepsilon$ and $\mu$ in a surrounding volume
$V_c$ to mimic a coordinate transformation, with Jacobian $\J$,
mapping the physical space $X$ to a virtual space $X'$ in which the
object is absent ($\J_{ij} = \partial x_i/\partial x_j'$), as shown in
\figref{cloak-transf}.  This is achieved by $\varepsilon =
\varepsilon_a \J \J^T / \det \J$ and $\mu = \mu_a \J \J^T / \det \J$
(for isotropic $\varepsilon_a$, $\mu_a$)~\cite{Pendry06}. (The surface
of the object in $X$ is mapped to the ground plane in $X'$, and so the
inner surface of the cloak must be reflective like the ground plane.)
We derive the limitations of cloaking under the following two
practical requirements:
\begin{itemize}
\item The attainable refractive index contrast
  $\sqrt{\varepsilon\mu}/\sqrt{\varepsilon_a\mu_a}$ (the eigenvalues of $\J
  \J^T / \det \J$) is bounded above by $B$ and below by $b$.
\item The scattering cross-section (nonzero due to imperfections in
  the cloak) for any incident wave is bounded above by some fraction
  $f$ of the geometric cross-section $s_g$.
\end{itemize}
Given these two constraints, we derive the following relations between
the difficulty of cloaking and the size of the object to be cloaked:
\begin{itemize}
\item The thickness of the cloak must scale with the object thickness
  (divided by $B$).
\item The allowed imperfections (e.g. disorder or absorption) must scale
  at most inversely with the object thickness.
\end{itemize}

\subsection{Cloak thickness} \label{sec:thickness}

\begin{figure}[t]                                           
\centering        
\includegraphics[width=0.5\textwidth]{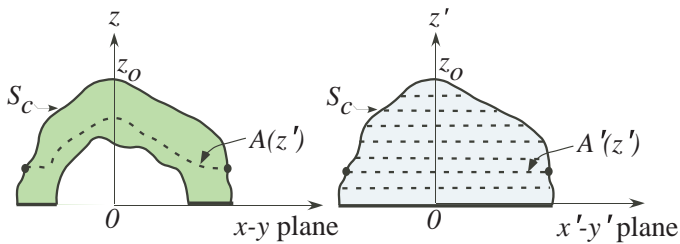}
\caption{The cloaked volume $V_c'$ in virtual space can be divided
  into flat cross-sections $A'(z')$ for each $z' \in [0,z_0]$. These
  are mapped to curved surfaces $A(z')$ in $X$. The invariance of the
  outer surface $S_c$ means that the boundaries (solid dots) of
  $A(z')$ and $A'(z')$ coincide, and hence $A(z')\geq A'(z')$.}
\label{fig:cloak-int} 
\end{figure}

The volume $V_c$ is given by $\int_{V_c'} |\det \J| dx'dy'dz'$, and
therefore constraints on $V_c/V_c'$ immediately follow from two
facts. First, $|\det \J|$ can be bounded due to the bound $B$ on the
index change above.  Second, an even tighter bound follows from
the fact that the outer surface $S_c$ of the cloak is invariant under the
coordinate transformation.  In particular, defining $\J_x$ and $\J_y$
as the first two columns of $\J$ and denoting singular values of $\J$
by $\sigma_i$ ($\sigma_i^2$ are eigenvalues of $\J \J^T$~\cite{Trefethen97}),
referring to the cross sections $A'(z')$ defined in
\figref{cloak-int}, we show the following sequence of bounds:
\begin{align}
  V_c &= \iiint_{V_c'} |\det \J| \, dx'dy'dz' \label{eq:Vc1} \\
      &\geq (\min \sigma_i) \int_0^{z_0} dz' \iint_{A'(z')} |\J_x \times \J_y| \, dx' dy' \label{eq:Vc2} \\
      &= (\min \sigma_i) \int_0^{z_0} A(z') \, dz' \label{eq:Vc3} \\
      &\geq (\min \sigma_i) \int_0^{z_0} A'(z') \, dz' = (\min \sigma_i) V_c' \label{eq:Vc4} \\
      &\geq V_c' / B. \label{eq:Vc5}
\end{align}
Step~\eqnumref{Vc2} follows from $|\det \J| = \Vert \J^T (\J_x \times
\J_y) \Vert$, and $\Vert \J^T \vec{u} \Vert \geq (\min \sigma_i) \Vert
\vec{u} \Vert$ for any vector $\vec{u}$~\cite{Trefethen97}. The $x'y'$ integral in
line~\eqnumref{Vc2} is simply the area $A(z')$ that $A'(z')$ maps to,
and $A(z') \geq A'(z')$ because the outer boundary (solid dots in
\figref{cloak-int}) is identical in $A$ and $A'$ and the flat surface
$A'$ is the minimal area for this boundary.  Finally, step \eqref{Vc5}
stems from elementary properties of singular values: the eigenvalues of $\J
\J^T / |\det \J|$ are simply $b \leq \sigma_i ^2 /
\sigma_1\sigma_2\sigma_3 \leq B$~\cite{Trefethen97}, and algebraic manipulation of this inequality yields $B^{-1} \leq \sigma_i \leq b^{-1}$.  Thus, we
have shown that $V_c \geq V_c'/B$, but since the outer surface $S_c$
is invariant it follows that the thickness of the cloak must scale
proportional to the object thickness divided by $B$.

\subsection{Cloak losses}

We will analyze losses due to imperfections via perturbation theory:
we first obtain the fields in a perfect cloak ($\varepsilon$ and $\mu$
given exactly by the transformation law), and then consider the
lowest-order absorption or scattering in the presence of small
imperfections.  Suppose the object is illuminated by an incident
planewave with electric-field amplitude $E_0$, which means that the
total field (incident + reflected) in the ambient medium and in $V_c'$
has amplitude $\leq 2E_0$.  For a perfect cloak, the fields in the
cloak $V_c$ are simply given by $(\J^T)^{-1}$ multiplied by the
reflected planewave in virtual space $X'$~\cite{Pendry06}, and hence
the field amplitude $E$ in the cloak is $\leq 2E_0 / (\min \sigma_i)
\leq 2E_0 B$ (using the bounds on the singular values $\sigma_i$ from
above).  We must consider the worst-case losses for arbitrary incident
waves; since this is bounded below by the loss from any particular
incident wave, it is convenient to consider glancing-angle
$p$-polarized planewaves where the field is a constant $E_0$
everywhere in $V_c'$ and $E \geq E_0 b$ in the cloak (from the bound
on $\max \sigma_i$).

\subsubsection{Absorption}
\label{sec:absorption}

An absorption imperfection is a small deviation $\Delta\Im\varepsilon$ in
the imaginary part of $\varepsilon$ compared with $\varepsilon_a \J \J^T / \det
\J$.  (Similarly for $\mu$, but it suffices to consider electric
absorption here.) This gives a change $\frac{\omega}{2} \Re \int_{V_c}
\vec{E}^* \Delta\Im\varepsilon \vec{E}$ in the time-average absorbed
power at a frequency $\omega$~\cite{Jackson98}.  To lowest order in
$\Delta\varepsilon$, we can take $\vec{E}$ to be the field in the perfect
cloak, and suppose for simplicity that the absorption is isotropic
($\Delta\Im\varepsilon$ is a scalar), and therefore the worst-case change
in the absorbed power is $\geq \frac{\omega}{2} E_0^2 b^2 |\int_{V_c}
\Delta\Im\varepsilon|$.  Combined with our initial requirement on the scattering
cross-section, the change in the
absorbed power must be $\leq fs_g I_0$ where
$I_0=\frac{1}{2}E_0^2\sqrt{\varepsilon_a/\mu_a}$ is the incident intensity,
and we obtain the following bound:
\begin{equation}
\mean \Delta\Im\varepsilon \leq \frac{f \sqrt{\varepsilon_a/\mu_a}}{\omega b^2} \frac{s_g}{V_c} \leq \frac{f \sqrt{\varepsilon_a/\mu_a} B}{\omega b^2} \frac{s_g}{V_c'},
\label{eq:loss-necessary}
\end{equation}
(using the $V_c$ inequality above).  The ratio 
$s_g/V_c'$ scales as $1/\mathrm{thickness}$, so this means that the
mean $\Delta\Im\varepsilon$ scales inversely with the thickness of the
object to be cloaked.  (The $1/\omega$ dependence means that this can
be interpreted as a bound on the conductivity.)

\Eqref{loss-necessary} is a necessary condition on the loss, but is
too optimistic to be a sufficient condition.  For example, suppose
that the ambient medium is lossy, so that $\Delta \Im \varepsilon$ can
have either sign depending on whether the cloak is more or less lossy
than the ambient medium (or alternatively, $\Delta\Im\varepsilon < 0$
could come from gain). \Eqref{loss-necessary} is satisfied if the
more-lossy and less-lossy regions of the cloak average to zero, but in
fact this will not result in a zero scattering cross-section for
arbitrary incident waves.  For example, a narrow incident beam (rather
than a planewave) will interrogate the loss in some regions of the
cloak more than others, and even a planewave at a different angle will
create a standing-wave pattern that has higher field intensity in some
regions---in these cases, any delicate cancellation
in the absorption will be destroyed.  Instead,
we can derive a sufficient condition on the loss by bounding the
change in absorption above rather than below.  In particular,
$|\frac{\omega}{2} \Re \int_{V_c} \vec{E}^* \Delta\Im\varepsilon \vec{E}
| \leq \frac{\omega}{2} V_c (2E_0 B)^2 \max |\Delta\Im\varepsilon|$, and
thus it is sufficient for
\begin{equation}
\max |\Delta\Im\varepsilon| \leq \frac{f \sqrt{\varepsilon_a/\mu_a}}{4\omega
  B^2} \frac{s_g}{V_c} \leq \frac{f
  \sqrt{\varepsilon_a/\mu_a}}{4\omega B} \frac{s_g}{V_c'} .
\label{eq:loss-sufficient}
\end{equation}
This is, perhaps, stronger than strictly necessary; we conjecture that
a weaker sufficient condition exists that replaces $\max
|\Delta\Im\varepsilon|$ by an average of $\Delta\Im\varepsilon$ in the
smallest region that can be interrogated by an incident wave
(i.e. some wavelength-scale region).  Regardless, both the sufficient
and necessary conditions on the absorption imperfections scale
inversely with the object thickness.  This is true regardless of
whether the ambient medium is lossy, and also means that any gain-based
compensation of absorption must become increasingly exact for larger
objects.

\subsubsection{Random imperfections}

Small random imperfections can be thought of as scatterers distributed
randomly throughout $V_c$ with some polarizability $\alpha$ (dipole
moment $\vec{p}=\alpha\vec{E}$)~\cite{JohnsonPo05}.  For example, a
small change $\Delta\varepsilon$ in a small region $\delta V$ corresponds
to $\alpha=\Delta\varepsilon \delta V$. Computing $\alpha$ for
surface roughness is more involved, but is conceptually
similar~\cite{JohnsonPo05}. If these imperfections are uncorrelated, then on
average the scattered power is simply the mean dipole radiation from
each scatterer multiplied by the density $d_\alpha$ of scatterers (the
radiation from different scatterers is incoherent, so interference
terms average to zero)~\cite{Jackson98}.  This radiation is most
easily computed by transforming each scatterer back to virtual space
$X'$, where the polarizability is $\alpha' = (\J^T)^{-1} \alpha
\J^{-1} \det\J$, so that $|\alpha'|\geq|\alpha|/B$ from above.  The
radiated power of a point source in virtual space (homogeneous above a
ground plane) varies with distance and orientation above the ground
plane due to the image dipole source below the ground plane, but the
worst-case (over all incident waves) average (over all scatterer
positions) scattered power is proportional (with a constant factor of
order unity) to the radiated power of a point source in the
homogeneous medium, $(\alpha' E_0)^2 \omega^4 \mu_a
\sqrt{\varepsilon_a\mu_a}/ 12\pi$~\cite{Jackson98}.  Multiplying this by
the number $d_\alpha V_c$ of scatterers and comparing to the
requirement on the worst-case loss, one finds that $\alpha^2 d_\alpha$
is bounded above by a quantity proportional to $B^3 f s_g/V_c'$, which
again scales inversely with the thickness of the cloaked object.

Note that, while gain could conceivably be used to compensate for
absorption loss, it does not seem applicable to scattering from
imperfections. 

\section{Cloaking of isolated objects}

\begin{figure}[t]                                              
\centering                                                     
\includegraphics[width=0.5\textwidth]{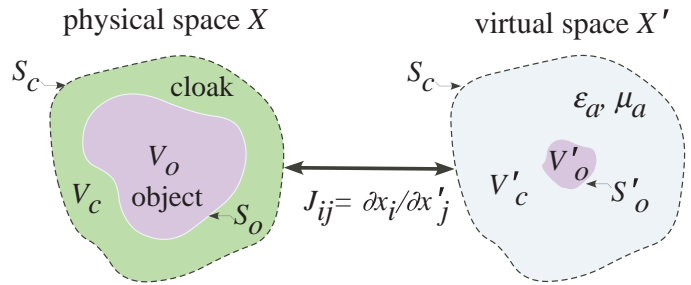}               
\caption{isolated-object cloak}                                
\label{fig:cloak-iso}                                          
\end{figure}   

Consider the problem of cloaking an isolated object of volume $V_o$ in
a homogeneous isotropic ambient medium using a transformation-based
cloak of volume $V_c$ surrounding the object. As above, the cloak
material is determined from a coordinate mapping with Jacobian $\J$,
via the equations $\varepsilon=\varepsilon_a\J\J^T/\det \J$ and $\mu =
\mu_a \J\J^T/\det\J$.  Now, however, there is no ground plane, and so
the coordinate mapping instead attempts to shrink the object: it maps
the physical space $X$ to a virtual space $X'$ in which the object
volume $V_o$is mapped to a smaller volume $V_o'$, as shown in
\figref{cloak-iso}.  (As for ground-plane cloaking, the outer cloak
surface $S_c$ is invariant, since the transformation is the identity
outside of the cloak.) Perfect cloaking corresponds to the case in
which $V_o'$ is a single point, but we will show that this is not
possible if the index contrast (eigenvalues of $\J \J^T/\det\J$) is
bounded below by $b > 0$ as above.  We will also show results
analogous to our results for ground-plane cloaking: if the index
contrast is bounded above by $B$, then the thickness of the cloak must
scale with the object diameter, and correspondingly the losses (from
absorption or imperfection) in the cloak must decrease with the object
diameter.

Before developing our general results, however, we will begin with a
specific illustrative example which demonstrates these scalings: an
adaptation of Pendry's linear-scaling cloak
design~\cite{Pendry06} to a nonzero $V_o'$.

\subsection{Example: A spherical linear-scaling cloak}

Suppose that the cloaked object is a sphere of radius $R_1$ and the
cloak has outer radius $R_2$.  We shrink the object to a sphere of
radius $R_1'$ with the transformation $r' = R_1' + (r - R_1) (R_2 -
R_1') / (R_2 - R_1)$ in spherical coordinates.  This leads to the following transformed materials in the cloak region (applying the general spherical-coordinate version of the transformation~\cite{Pendry06}:
\begin{align}
\varepsilon_\theta/\varepsilon_a=&\mu_\theta/\mu_a= \varepsilon_\phi/\varepsilon_a = \mu_\phi/\mu_a =\frac{R_2-R_1'}{R_2-R_1}, \label{eq:r} \\
\varepsilon_r/\varepsilon_a=& \mu_r/\mu_a = \frac{R_2-R_1}{R_2-R_1'}\frac{\left[R_1' + \frac{r-R_1}{R_2-R_1}(R_2-R_1')\right]^2}{r^2} \label{eq:theta}
\end{align}

At the inner cloak surface $r=R_1$, the radial components of
$\varepsilon$ and $\mu$ simplify to $\frac{R_2 - R_1}{R_2 - R_1'}
(R_1' / R_1)^2$, which vanishes for $R_1'=0$.  If we impose a lower
bound $b$ on the singular values of $\varepsilon\mu /
\varepsilon_a\mu_a$, however, then $R_1'$ cannot vanish.  In
particular, if $R_1 \gg R_1'$ then $(R_1' / R_1)^2 \sim b$, and hence
the area reduction $S_o'/S_o \sim b$.  Below, we show in general (for
arbitrary non-spherical transformations) that a $b>0$ condition imposes a lower bound on the area reduction.

Even if $b=0$, there is still a relationship between the
index-contrast upper bound $B$ and the cloak thickness.  Suppose $R_1'
\ll R_1$ (i.e. it is an effective cloak).  Then the tangential
components of $\varepsilon$ and $\mu$ are $\approx R_2 / (R_2 - R_1)
\leq B$, from which it immediately follows that the cloak thickness
$R_2 - R_1$ is bounded by $R_2/B \sim R_1/B$.  This is identical to
what we found for ground-plane cloaking, above: cloak thickness scales
with object thickness divided by $B$.  Below, we will generalize this
thickness scaling to arbitrary non-spherical cloaks.

\subsection{General limits on cloaking cross section}

In this section, we show in general that the $S_o'$, the effective
surface area of the cloaked object, must be $\geq S_o b^2$.  That is,
if $b>0$, then the area $S_o'$ as seen by an observer must scale
proportional to the object area, so that the cross section is reduced
by a bounded factor.  (For objects much larger than the wavelength,
the scattering cross section is proportional to the geometric cross
section $S_o'$.)  From the spherical example above, it might be
possible to derive an even tighter $\sim b$ bound on $S_o'$, but the
$S_o b^2$ bound here is sufficient to demonstrate the scaling with
$S_o$.

Our basic approach is to write down $S_o$ in terms of an integral over
$S_o'$, and then to bound the integral:
\begin{equation}
S_o = \oiint_{S_o'} \left|\J_u \times \J_v\right| dA'(u',v'),
\end{equation}
where $(u,v) \leftrightarrow (u',v')$ is some coordinate system of the
surfaces, $dA'(u',v')$ is the area element in $S_o'$, and $\J_u$ and
$\J_v$ are columns of the Jacobian matrix as in \secref{thickness}.
We must then bound $|\J_u \times \J_v|$ similar to our previous
analysis, but this is conceptually complicated somewhat by the fact
that the coordinate system here is non-Cartesian, meaning that $\J$ no
longer has the same bounds.  To circumvent this difficulty, we apply a
standard trick from differential geometry~\cite{Carmo92}: we cover the surface
$S_o'$ with small overlapping neighborhoods $N_k'$ (a locally finite
open covering~\cite{Carmo92}) where the surface is locally approximately flat,
in which case we can define a local Cartesian coordinate system and
use the Cartesian bounds on $\J$ (ultimately taking the limit of
infinitesimal neighborhoods so that the local flatness becomes exact).  To
combine these local results, one uses a partition of unity~\cite{Carmo92}: a
set of ``bump'' functions $p_k'(u',v')$ (nonzero only on $N_k'$) such
that $\sum_k p_k' = 1$ on $S_o'$.  Thus, we obtain:
\begin{equation}
S_o = \sum_k \iint_{N_k'} \left|\J_u \times \J_v\right| p_k' dA'.
\end{equation}
Now, in the limit of infinitesimal neighborhoods $N_k'$, we can freely
treat each integral as being over a local Cartesian coordinate
system, in which case $\J$ is the ordinary Cartesian Jacobian matrix.
Now, we use two facts.  First, we know $|\det \J| = \sigma_1 \sigma_2
\sigma_3 \leq (\max \sigma_i)^2 (\min \sigma_i)$. Second, similar to
\secref{thickness}, $|\det \J| = \Vert \J^T (\J_u\times\J_v)\Vert \geq
(\min \sigma_i) |\J_u\times\J_v|$ from general properties of singular
values~\cite{Trefethen97}.  Combining these two inequalities, we find
$|\J_u\times\J_v| \leq (\max \sigma_i)^2 = 1/b^2$.  Substituting this bound into the integral above, we finally obtain:
\begin{align}
S_o &\leq \sum_k \iint_{N_k'} p_k' dA' / b^2 \\
    &= \oiint_{S_o'} \left(\sum_k p_k'\right) dA'/b^2 = \oiint_{S_o'} (1) dA'/b^2 \\
    &= S_o'/b^2,
\end{align}
the desired inequality.  (Note that if the object contains corners
where $S_o'$ is not locally flat, that does not affect this analysis
since those corner regions have zero measure in the integral.
The bounds on $\J$ mean that corners in $S_o$ must be mapped
to corners in $S_o'$ and vice versa, and the integrand is finite.)

We suspect that a tighter bound, proportional to $b$ instead of
$b^2$, can be proved by taking into account the fact that the
coordinate transformation must leave $S_c$ invariant.  In the
spherically symmetrical example above, the purely radial nature of the
coordinate transformation caused it to have at most one factor of
$1/b$ in $|\det \J|$ (only one eigenvalue of $\varepsilon$ and $\mu$
vanishes for $R_1'=0$) and not two, leading to an $S_o \sim S_o'/b$
dependence.  A similar single factor of $b$ in the general case would
give a $1/b$ scaling in the inequalities above by replacing $(\max
\sigma_i)^2$ with $\max \sigma_i$.

\subsection{General scaling of cloak thickness and loss}

A simple linear scaling of the cloak volume, as mentioned at the beginning of \secref{thickness}, follows immediately from the bounds on $|\det \J|$:
\begin{equation}
V_c= \iiint_{V_c'} |\det \J| \, dx'dy'dz'\geq (\min \sigma_i)^3V'_c \geq V'_c/B^3
\label{eq:iso-vol}
\end{equation}
As for ground-plane cloaking, we can define a mean thickness $V_c/S_c$
of the cloak.  For a useful cloak, $V_o' \ll V_o$ and hence $V_c'
\approx V_c + V_o$.  Thus, $V_c'/S_c' = V_c'/S_c$ is a mean total
(cloak + object) diameter.  Therefore, we have just demonstrated the
inequality $V_c/S_c \geq (V_c'/S_c')/B^3$ (to lowest order in
$V_o'/V_o$), which means that the mean cloak thickness $V_c/S_c$ must
scale proportional to the object diameter.

The spherical Pendry example above leads us to suspect that a tighter
bound $\sim 1/B$ can be derived.  Similar to the ground-plane example
in \secref{thickness}, we expect that the key point is that we have
not yet taken into account the constraint $S_c' = S_c$ on the cloaking
transformation.  However, our main goal here is to demonstrate the
scaling of cloak thickness with object thickness, not to fine-tune the
constant factor.

Now that we know that cloak thickness must scale with object
thickness, an analysis of losses similar to that in the ground-plane
case above must apply.  As the object becomes thicker, incident rays
travel for a longer distance through the cloak, and hence for a fixed
tolerance on the cross section the losses per unit distance must
shrink proportional to the object diameter.  Hence absorption losses
(or rather, the \emph{difference} between the cloak absorption in $X'$
and the absorption of the ambient medium) and imperfections must scale
inversely with the diameter.  As we pointed out in our previous
paper~\cite{Hashemi10}, precisely such an inverse scaling of absorption
tolerance with diameter has been demonstrated numerically~\cite{Baile}
for spherical cloaks, and our work now shows that this relationship is
general.  As suggested by some authors~\cite{Han11}, gain could
be used to compensate for absorption (but not disorder), but as
discussed above our results imply that this compensation must become
more and more exact as the object diameter increases.

For example, let us explicitly consider absorption imperfections for
the idealized case of $b=0$: that is, suppose we are able to map $V_o$
to a single point $V_o'=0$, but have a finite $B$ and are still
concerned with imperfect materials.  In this case, an incident
planewave of amplitude $E_0$ corresponds to a planewave of constant
amplitude in the cloak for perfect materials, or to lowest order in
the imperfections for imperfect materials.  Then, similar to our
analysis in \secref{absorption}, the change in absorbed power is
bounded below by $\frac{\omega}{2} E_0^2 |\int_{V_c}
\Delta\Im\varepsilon / (\max \sigma_i)^2|$.  Since we require that
  this change also be bounded above by the incident intensity ($\sim
  E_0^2$) multiplied by some fraction $f$ of the geometric
  cross-section $s_g$, we obtain a necessary condition on the
  absorption imperfection as in \secref{absorption}.  In particular it
  follows that an averaged absorption imperfection $|\int_{V_c}
  \Delta\Im\varepsilon / (\max \sigma_i)^2|/V_c$ must scale
    proportional to $s_g/V_c \sim s_g B^3 / V_c'$, where $V_c'/s_g$ is
    proportional to the diameter.  In \secref{absorption}, we further
    used $\max \sigma_i \leq 1/b$ and pulled out a $b^2$ factor, but
    that is not appropriate when $b=0$, so instead we must leave a
    $1/(\max \sigma_i)^2$ weight factor (which is only zero at the
    inner surface of the cloak) in the average of
    $\Delta\Im\varepsilon$.  As in \secref{absorption}, this is not a
    sufficient condition because the observer need not use
    planewaves---for interrogating the cloak with a focused beam, a
    stronger condition must apply, in which $\Delta\Im\varepsilon$
    within a small (wavelength-scale) volume must go to zero as
    diameter increases.  If $b>0$, this analysis is only slightly
    modified in principle (although the precise expression becomes
    much more complicated): the small scattered field (assuming $V_o'\ll
    V_o$) from the nonzero $V_o'>0$ modifies the field in $X'$ by a
    small amount over most of $V_c'$ (except immediately adjacent to
    $V_o'$), which should only change the proportionality of the $\Im
    \Delta\varepsilon$ scaling by a small factor.

\section{Conclusions}

Generalizing our previous work~\cite{Hashemi10}, these scaling laws
point to an inherent practical difficulty (though not a mathematical
impossibility) in scaling experimental cloaking of small objects to
larger ones.  Furthermore, we showed that very similar analysis can be
applied to cloaking of isolated objects---bounded index contrasts will
imply a bounded reduction in the scattering cross section, a cloak
thickness proportional to the object diameter, and imperfection
tolerances that shrink with the object diameter (a scaling we already
observed numerically~\cite{Hashemi10}).  It might be possible to
further generalize the results in this paper to cloaks that are not
derived from coordinate transformations (similar to the generality of
our one-dimensional analysis~\cite{Hashemi10}).  However, the most
serious constraint on isolated-object cloaking seems to be the
bandwidth, which must be zero for perfect cloaking~\cite{Pendry06,
  Miller06}.  Clearly, if perfect cloaking is possible (theoretically)
at a single frequency, then imperfect cloaking (reduction of the cross
section by a given factor) must persist over a nonzero bandwidth, and
an interesting open problem is to prove how the bandwidth of such
imperfect isolated-object cloaking must scale with the object diameter
from causality constraints.

An alternative direction is to consider relaxations of the cloaking
problem that might prove more practical.  In particular, it would be
valuable to make precise the intuition that the cloaking problem
becomes easier if the incident waves are restricted (e.g. to
planewaves from a certain range of angles) and/or the observer is
limited (e.g. only scattered waves at certain angles are visible, or
only amplitude but not phase can be detected), since this is arguably
the situation in most experiments. (For example, current ``stealth''
aircraft are designed in the radar regime mainly to reduce
back-scattering only~\cite{Nicolai10}.)  Another interesting
possibility is to consider ``cloaking'' that attempts to make one
object look like a different object of a similar size rather than
making it invisible (although this approach is similar in spirit to
ground-plane cloaking and may have similar limitations).

This work was supported in part by the Army Research Office through
the Institute for Soldier Nanotechnologies (ISN) under contract
W911NF-07-D-0004, and by the AFOSR Multidisciplinary Research Program of the University Research Initiative (MURI) for Complex and Robust
On-chip Nanophotonics, Grant No. FA9550-09-1-0704.


\end{document}